# Derivation of the Probability Law in the Many-Worlds, One-MIND Interpretation

Casey Blood
Rutgers University, retired
Sarasota, FL
CaseyBlood@gmail.com

## Abstract

The basic mathematical structure, QM-A, of the many worlds interpretation consists solely of the linear mathematics plus the Hilbert space properties of the state vectors. There is no collapse and there are no particles or hidden variables. It is remarkable that QM-A alone can account for all our observations except probability. There is no need for particles, hidden variables or collapse to explain perception of only one classical version of reality, the photoelectric effect, localized effects from a spread-out wave function in scattering and interference experiments, wave-particle duality, and so on. But probability cannot be defined within QM-A.

Nevertheless, because of its astonishing success, it seems reasonable to require (1) that the mathematics of an interpretation be limited to the highly successful QM-A and (2) that "matter" be composed of state vectors alone. But the probability law requires, in essence, that one version of the observer be singled out. The most straightforward way to accomplish this under (1) and (2) is to assume there is an aspect of the observer—the Mind—which is outside the laws of quantum mechanics and perceives just one version of reality. Under that assumption, the $|a(k)|^2$ probability law can then be derived. Thus we have an interpretation, QM-A plus "outside" observer, which explains all our perceptions.

## 1. Introduction.

Quantum mechanics is an extremely successful theory, giving accurate predictions in elementary particles, atomic and molecular systems, and many-body physics. And it has never predicted a result in conflict with experiment. In spite of its amazing success, however, quantum mechanics has one most peculiar property; it gives several simultaneously existing versions of reality. For example, Schrödinger's cat is both alive and dead at the same time. Quantum mechanics implies we will perceive only one of those versions of reality (Sec. 2.3), but it cannot predict which one.

However, if an experiment is run many times, quantum mechanics can predict the *probability* of perceiving a given outcome. The probability law, Born's rule, does not follow, at least not directly, from the equations of motion for the state vectors and so at first glance it appears the law must be assumed separately from those equations.

Zurek [1,2] has given a derivation of the law but, as we will discuss, it is based on questionable assumptions. We propose a different derivation here, based on the freedom of choice of basis, the normalization condition, and the assumption that the concept of probability is indeed relevant to the conceptual scheme used.

Our starting point is the basic mathematics of quantum mechanics—the linear equations of motion and the Hilbert space of the state vectors. There is no collapse, there are no hidden variables, and nothing (such as particles) exists besides the state vectors. We call this scheme, which is essentially the same as the starting point of Everett's many-worlds interpretation [3], QM-A. It is remarkable that all the characteristics of our perceptions— localized detection of "particles" from a spread-out wave function in scattering and interference experiments, perception of a single version of reality, the photoelectric and Compton effects, the particle-like properties of mass, energy, momentum, spin and charge, wave-particle duality, long-range correlations in Bell-type experiments, the Heisenberg uncertainty principle—all these can be shown to follow from the mathematics of QM-A. *There is no need to assume either collapse or the existence of particles or hidden variables to explain all these phenomena.*

However, there is still a significant problem because there is manifestly no *probability* of perception within QM-A, for every outcome is perceived (by a version of the observer) on every run. Thus the "pure" QM-A does not give an acceptable interpretation of quantum mechanics.

To account for probability, a number of interpretations—which typically take the form of mathematical amendments to QM-A—have been proposed, the most prominent being collapse of the wave function and hidden variables. There is, however, no evidence for any of these interpretations, there is significant experimental evidence against several of them, and there are theoretical hurdles to overcome in constructing them. Further, aside from Everett's many worlds, none of the currently proposed interpretations fully acknowledge the near-perfect success of QM-A.

This leaves us with an interesting choice. We can hope that eventually hidden variables, or collapse, or some other interpretation in which the QM-A mathematics is amended will be shown to be correct—capable of explaining all our perceptions, including the probability law. *Or*, because QM-A does so well, we can explore interpretations in which the mathematics consists solely of the linear equation, Hilbert space structure of QM-A. We choose the latter here. In that case, the probability law virtually forces us to the assumption that the observer, in addition to being associated with a brain-body wave function/state vector, also has an aspect, outside the mathematics of quantum mechanics, which is aware of just one version of reality. The most basic ramifications of that assumption are considered here.

In Sec. 2, we outline the steps necessary to show that QM-A can account for all perceptions except the probability law. In Sec. 3, we sketch the problems associated with hidden variable and collapse interpretations. Then in Sec. 4, we derive the probability law, under the assumption that probability of perception is consistent with the scheme. The conflict between QM-A and probability is discussed in Sec. 5.

We then introduce in Sec. 6 the idea that there is, associated with each individual, a non-quantum mechanical Mind, an aspect of the observer which is outside the laws of quantum mechanics. This Mind, presumed to perceive just one version of the brain-body of the observer, allows us to properly introduce the concept of probability. But to make the scheme consistent, we must also suppose there is an overarching MIND, with each individual Mind being a part or facet of the MIND. The questionable aspects of Zurek's derivation of the probability law are discussed in Sec. 7. And finally this many-worlds, one-MIND interpretation, m-w*, consisting of QM-A plus the non-quantum mechanical, perceiving aspects of the observers, is evaluated in Sec. 8.

## 2. The Strengths of QM-A.

We wish to see what the pure (free of add-ons such as collapse or hidden variables) mathematics of QM-A, by itself, implies about our perceptions of the physical universe. We find QM-A explains everything [4,5] except the probability law, so there is very little left for interpretations or modifications of quantum mechanics to explain.

### 1. The state vector.

We start by reviewing the state vector for an experiment, with the prototypical example being the Stern-Gerlach experiment. The original atomic state, just before detection, is

$$|\Psi, t_0\rangle = \sum_{k=1}^{n} a(k)|k\rangle \qquad (1)$$

The states $|k\rangle$ are defined by the fact that they are the part of $|\Psi\rangle$ which travels along the $k^{th}$ "trajectory." These states are determined by the original state vector plus the interaction Hamiltonian between the "particle" (particle-like state vector) and the device (magnet) which separates the state into different "trajectories." Thus the $|k\rangle$ and the $a(k)$ are independent of the basis used to express the particle state before it passes the device (although the *expression* of the state will be different in different bases). That is, no specific basis has been chosen in writing Eq. (1) (but the $|k\rangle$ are often chosen as basis vectors).

We suppose the wave function for each possible outcome is detected by a separate detector. The particle-detector Hamiltonian is such that atomic state $|k\rangle$ interacts only with detector $|d:k\rangle$, so that $|k\rangle$ changes the reading on $|d:k\rangle$ from "no" (no detection) to "yes" (detection). We abbreviate the product detector state in which only detector $k$ reads yes while the all other $n$-1 detectors read no by $|D:k,\text{yes}\rangle$. Thus after "the particle" has hit the detectors, the state vector is

$$|\Psi, t_1\rangle = \sum_{k=1}^{n} a(k)|k\rangle \, |D:k,\text{yes}\rangle \qquad (2)$$

Finally we suppose there is an observer, with the detector-observer Hamiltonian being such that the detector state $|D:k,\text{yes}\rangle$ causes the observer's neurons to fire in a pattern indicating state $k$ has been detected. The state is then

$$|\Psi, t_2\rangle = \sum_{k=1}^{n} a(k)|k\rangle \, |D:k,\text{yes}\rangle \, |\text{Obs perceives } k\rangle \qquad (3)$$
$$= \sum_{k=1}^{n} a(k)|\text{branch } k\rangle$$

Thus at the end there are *n* different branches to the state vector, each corresponding to a different outcome, with each branch containing a different version of the observer (that is, there is no singular "the" observer).

Note that there has been no assumption of a specific basis at any of the three stages; rather the initial atomic state, the orientation of the Stern-Gerlach apparatus, and the interaction Hamitonians determine the basis-independent states at each stage.

## 2. Orthogonality of, and non-communication between, branches.

First, because the versions of the detectors are macroscopically different on the different branches, we can safely assume that the state vectors of the versions are mutually orthogonal,

$$\langle D{:}j,\text{yes}|D{:}k,\text{yes}\rangle = \delta_{jk} \qquad (4)$$

Further, no relevant time evolution will change the readings on the detectors, so we also have, at time *t*,

$$\langle D(t){:}j,\text{yes}|D(t){:}k,\text{yes}\rangle = \delta_{jk} \qquad (5)$$

Now suppose we consider the matrix elements of the system Hamiltonian between the detector states at time t. The Hamiltonian does not, by hypothesis, change the readings on the detectors so, because the states of the atoms making up the detectors are orthogonal for different readings (for example, a pointer pointing in different directions), we have

$$\langle D(t){:}j,\text{yes}|H|D(t){:}k,\text{yes}\rangle = 0, j \neq k \qquad (6)$$

and therefore

$$\langle \text{branch } j(t)|H|\text{branch } k(t)\rangle = 0, j \neq k \qquad (7)$$

This implies there can be no interaction between branches. For example, a photon emitted by the *k* version of the detector must stay paired with the *k* version of the detector; Eq. (7) implies the photon cannot migrate to the *j* branch (where it would be paired with the *j* version of the detector). Thus Eq. (7) implies the photon cannot interact with—be perceived by—version *j* of the observer, who is, of course, always on the *j* branch. The result is that version *j* of the observer can perceive only what happens on its own branch. There can be no passing of information, no communication, between branches.

## 3. No non-classical perceptions.

First note in Eq. (3) that the events on each branch are "if-then classical:" if the atomic state is $|j\rangle$, then the detectors read *j*; and if the detectors read *j*, then the observer perceives *j*. Further since version *j* of the observer can perceive what happens only on its own branch, no version of the observer perceives what happens on more than one branch (a "double-exposure"). This implies QM-A never leads to a non-classical perception. Note that since no basis was chosen in writing Eq. (3), these results are guaranteed by the structure of the Hamiltonians, regardless of the basis chosen. This general argument would seem to be sufficient to show that QM-A alone leads to classical perceptions.

However, since the double-exposure question is contentious (in the form of the "preferred basis" problem), we will elaborate upon this point. The idea is to show that QM-A implies we never perceive more than one version of reality. But to do this, we need to note a property of our perceptions. If we ignore fleeting and indistinct perceptions, they are always *communicable*. That is, we can write down the results of a measurement.

> *Non-communicable perceptions are therefore of no relevance in the comparison of QM-A with our actual perceptions.*

And so our objective is to show that versions of the observer can never *communicably* perceive more than one version of reality in QM-A (with each version being classically consistent). This is sufficient to show a match between QM-A and our *communicable* perceptions. And since all our perceptions are communicable, this shows a match between QM-A and all our perceptions of physical reality.

We start by instructing the observer, before the experiment begins, to write down (communicate) what result he perceives—writing $j$ if a classical version $j$ is perceived and $\infty$ if any type of non-classical result is perceived. Then, using the time evolution of the brain-body of each (isolated) version of the observer to get from the $t_2$ of Eq. (3) to the state after writing, we obtain

$$|\Psi, t_3\rangle = \sum_{k=1}^{n} a(k)|k\rangle \, |D{:}k,\text{yes}\rangle \, |\text{Obs perceives } k, \text{writes } k\rangle \qquad (8)$$

Note that $\infty$ is never written. Further, no particular basis was used in writing Eq. (8); the state structure follows from the Hamiltonians alone. Thus $\infty$ will never be written in any basis.

To see this in more detail, consider a two-state system and suppose that the states potentially corresponding to our perceptions at time $t_2$, after detection and observation but before writing, are not $|\text{Obs } 1\rangle$ and $|\text{Obs } 2\rangle$ but rather are

$$|\text{Obs } a(t_2)\rangle = (|\text{Obs } 1(t_2) \text{ sees } 1\rangle + |\text{Obs } 2(t_2) \text{ sees } 2\rangle)/\sqrt{2}, \qquad (9)$$
$$|\text{Obs } b(t_2)\rangle = (|\text{Obs } 1(t_2) \text{ sees } 1\rangle - |\text{Obs } 2(t_2) \text{ sees } 2\rangle)/\sqrt{2}$$

Now the time evolutions of states |Obs 1> and |Obs 2> are fixed (by the memories of states 1 and 2 at $t_2$ and the brain dynamics leading to writing). So as time goes from $t_2$ to $t_3$, after writing, the $a$ and $b$ observer states evolve to

$$|\text{Obs } a,(t_3)\rangle = (|\text{Obs } 1(t_3) \text{ sees } 1, \text{writes } 1\rangle + |\text{Obs } 2(t_3) \text{ sees } 2, \text{writes } 2\rangle)/\sqrt{2}, \qquad (10)$$
$$|\text{Obs } b,(t_3)\rangle = (|\text{Obs } 1(t_3) \text{ sees } 1, \text{writes } 1\rangle - |\text{Obs } 2(t_3) \text{ sees } 2, \text{writes } 2\rangle)/\sqrt{2}$$

Thus we see that even though we have chosen a basis with "mixed" observer states, ∞ is still never written. That is, no matter what basis is used for the observer states, a double-exposure state, communicated by writing ∞, will never be communicably perceived by any version of the observer.

One might argue that Eq. (10) implies observer version *a* "obviously" perceives both results 1 and 2. But the situation requires a more detailed analysis. All our human perceptions of physical realty are communicable. But QM-A implies all the communicable perceptions of versions of the observer contain one and only one version of reality; ∞ *is not written in Eq.* (10). This agrees with our perceptions, which can be summarized as: *Observers never communicably perceive anything other than a single, classical version of reality.* Thus, although mathematically permitted, Eq. (9) is irrelevant for our perceptions because the "1" aspect of the brain of observer *a* does not have access to what the "2" aspect perceives (so both versions of reality cannot be communicably perceived at once).

**Second argument, using light.** There is another way to illustrate that version *a* of the observer does not perceive, in any relevant sense, a double exposure. We suppose that if the detector on path 1 registers yes, a blue light is shone into the observer's eyes, and if the detector on path 2 registers yes, a yellow light is shone. By using mirrors, it is arranged so that the two beams of light travel along the same path to the eyes, so if both lights are on, the observer sees both blue and yellow, which is perceived as green. Thus green signifies a non-classical perception (both 1 and 2 are perceived). But version *a* of the observer will never see green because version 1 perceives blue and, *in a separate universe*, version 2 sees yellow. By any reasonable definition of perception, this implies no double exposure is perceived.

In summary, *nothing is ever perceived by the versions of the observer in QM-A which is in conflict with our perceptions.*

Note that there is no need to use either the methods of decoherence [6] or the method of Zurek [1,2]— which employs **quantum Darwinism, Shannon entropy, and multiple records in the environment—to** show there is never (communicable) perception of more than one branch. Orthogonality of the detector states alone is sufficient.

## 4. The Photoelectric and Compton Effects.

The photoelectric and Compton effects are perhaps the primary evidence in attempting to show the existence of particles. Classically, light was considered to be a wave. But the photoelectric effect, in which electrons are ejected from a metal surface by shining light on it, was very difficult to explain using classical ideas. The problem was that the electromagnetic wave was spread out over many billions of electrons in the metal, so each individual electron received (classically) only a very small amount of energy per second. In fact, using classical ideas, it should have taken days for an electron to gain enough energy to be ejected from the metal. Experimentally, the ejected electron current started almost immediately.

Einstein proposed that this could be understood if one assumed there was a localized particle, a photon, which was embedded within the light wave function and carried all the energy. This idea, using the Bohr formula $E = h\nu$ for the photon energy, was sufficient to account for all the data.

The photon idea was seemingly confirmed in Compton scattering. If one assumes (1) there is a particulate electron with relativistic energy $E = \sqrt{m^2c^4 + p_e^2\,c^2}$, (2) there is a particulate photon that carried energy and momentum $E = h\nu, p = h/\lambda$, and (3) that energy and momentum are conserved in a collision, then the correct equations describing Compton scattering can be derived.

Does this prove there are particulate photons and electrons? No it doesn't, because one can also derive the photoelectric and Compton effect formulas using only properties of the wave function, with no assumption that particles exist [4,5]. There are two parts to the derivation. The first is to argue from group representation theory that mass, energy, momentum, spin and charge are actually properties of the wave function. And the second is to argue that, in contrast to the classical properties of waves, a small part of a spread-out wave function can transfer the full complement of energy and momentum to another, localized wave function.

## 5. Localization Effects.

Another seemingly strong piece of evidence for particles pertains to the localized effects of spread-out wave functions. For example, if a photon-like wave function goes through a single slit, becomes spread out and hits a screen covered with grains of film, one will find only a single grain of film exposed. Or if an electron-like wave function is scattered off a proton (wave function) so there is an outward spherical wave, and the detector is a sphere covered with film grains, again only a single grain of film will be exposed even though the wave function hits all the grains.

It is tempting to interpret these results as implying that there is a localized particulate photon or electron, embedded in the wave function, which hits and exposes just one grain. However, one can show (see [4,5]), using the results of Sec. 2.3 and the structure of the particle-detector Hamiltonians, that quantum mechanics, by itself, leads to the *perception* of only one localized grain exposed. And so the perception of localized effects from a spread-out wave function does not provide evidence for the existence of particles (or for collapse, or hidden variables) because it can be explained by QM-A alone.

One can extend this argument to show that QM-A, by itself, also predicts we will perceive particle-like trajectories in cloud and bubble chambers.

## 6. The particle-like properties of mass, energy, momentum, spin and charge.

In classical physics, it is assumed that particles possess the properties of mass, energy, momentum, spin (angular momentum), and charge. So if we wish to show there is no evidence for particles, we must show that these particle-like properties can be logically attributed instead to the wave functions/state vectors.

This is done using group representation theory [4,5]. The equations for the state vectors are linear and invariant under inhomogeneous Lorentz transformations (four-dimensional "rotations" plus translations) and internal symmetry group operations [7,8]. Invariance under inhomogeneous Lorentz transformations implies that the solutions— the state vectors—can be labeled by mass, energy, momentum, and angular momentum and its z component [9]. In addition, invariance under internal symmetry groups implies the charges are also properties of the state vectors.

So we see that the equations of quantum mechanics, using only the very general principles of linearity and invariance, rigorously imply that mass, energy, momentum, spin, and charge are properties of the state vectors. The consequence is that these properties—which along with localization essentially *define* particles—can reasonably be attributed to the state vectors.

### 7. Other phenomena and concepts.

First, wave-particle duality. *There is no evidence for the existence of particles* because all the properties we attribute to particles can instead be explained as properties of the state vectors (see references [4,5], summarized in Secs. 2.3-2.6). That is, there is no evidence for a *particulate* electron, separate from the state vector, but there are *state vectors* which have spin ½, charge –*e*, and an associated mass equal to that which we measure for "an electron". Thus wave-particle duality is simply a duality in the wave-like (principally interference) and particle-like (mass, spin, charge, localized effects, particle-like trajectories) properties of the state vectors. This duality of *properties* does not imply that matter is some peculiar amalgam of particles and waves.

Next, long-range correlations. In Bell-like experiments [10-14], there are correlations between the properties of entangled "particles," even if they are separated by macroscopic distances. These correlations are predicted perfectly by (no-particle) QM-A. The trouble in understanding them comes when one assumes there are actual particles which carry certain properties, such as polarization. If there are no particles, if all phenomena can be understood by using the properties of the state vectors alone, all the inferred "mysteries" in these phenomena—for example, instantaneous action at a distance between particles; or decisions "after the fact"—are eliminated.

Finally, there is the uncertainty principle, which seems to say certain properties of "particles" cannot be simultaneously measured with arbitrary accuracy. Again, the "mystery"—Why can't you simultaneously measure the position and momentum of a particle?—comes from assuming there are particles, with certain properties (position, momentum). If there are no particles, then the uncertainty principle is just a mathematical theorem about properties of the wave function/state vector.

## 3. Problems with Collapse and Hidden Variable Interpretations.

## Collapse.

Before starting our evaluation of collapse theories, we note that there is essentially only one reason to hypothesize collapse, and that is as an explanation of the probability law. For as we showed in Sec. 2, QM-A alone can explain all our other perceptions of the physical world.

### Collapse implies a non-linear theory.

We first show that collapse cannot occur in a linear theory. The state at time 0, including the detectors, is

$$|\Psi(0)\rangle_m = \sum_{k=1}^{n} a(k)|k\rangle_m |D:k, yes\rangle_m \tag{11}$$

where the index *m* refers to the $m^{th}$ run of the experiment. Let T(t) be the linear but not necessarily unitary time translation operator. Because of the linearity, the state at time t can be written as

$$|\Psi(t)\rangle_m = T(t)|\Psi(0)\rangle_m = \sum_{k=1}^{n} a(k)T(t)|k\rangle_m |D:k, yes\rangle_m \tag{12}$$
$$= \sum_{k=1}^{n} a(k)\beta_m(k,t)|k,t\rangle_m |D:k, yes\rangle_m$$

where the $\beta_m(k,t)$ indicates that the normalization of the state may change because of the possible non-unitarity of T(t). In a collapse interpretation the quantity

$$X_m(k,t) = \frac{|\beta_m(k,t)|^2}{\sum_{j=1}^{n} |\beta_m(j,t)|^2} \tag{13}$$

goes either to 0 or to 1 on each run as *t* becomes sufficiently large. If the probability law is to be satisfied, it goes to 1 on a fraction $|a(k)|^2$ of the runs and 0 on all the other runs so that, for large N and t,

$$(1/N) \sum_{m=1}^{N} X_m(k,t) = |a(k)|^2 \tag{14}$$

But because of linearity and the definition of the $\beta$s in Eq. (12), the $\beta$s and therefore the *X*s do not depend on the coefficients *a(k)*. Thus Eq. (14) cannot hold in a linear theory, so that any collapse theory must be non-linear. This is a huge proposed alteration of quantum mechanics because linearity is its most important characteristic.

**The GRW-Pearle Model.**
The most highly developed mathematical model of collapse is that of Ghirardi, Rimini, and Weber, and Pearle [15-18]. In it, each particle (particle-like wave function) feels a very weak "force" that acts only to change the normalization constants *a(k)*. The forces fluctuate randomly and are different in different regions of space. This has the advantage that collapse does not occur if there are only a very few particles, so few-body interference effects are preserved. But collapse can occur when there is a macroscopic number of particles in different states on the different branches, such as when a version of a detector displays 0 on one branch and another version of the same detector displays 1 on a second branch.
The main problem specific to the GRW-Pearle model is that the fluctuating forces—which are different in macroscopically separated regions of space within a branch, *and* different in the different branches—are all *instantaneously coordinated* (so that they tend to make the overall normalization larger). This instantaneous intra-and inter-branch coordination seems like a scheme that is very unlikely to occur in nature. (Note that the theory must be non-linear to have coordination between branches.)

More generally, it seems difficult, not just in this model but in any model, to devise a mathematical theory of collapse without having fluctuations that are coordinated both within and between branches.

A second problem is that the origin of the forces which cause collapse is not known. There has been some speculation that it could be due either directly to fluctuations in the vacuum state, or indirectly to the effects of gravity [19] on the vacuum state. But no progress has been made in converting this suggestion into a reasonably specific mathematical model. In particular—because a theory which includes the vacuum state is still a linear theory in current quantum mechanics—a new, radically different, non-linear theory of quantum mechanics must be specified.

A third problem is that the explanation of the probability law is just moved back one notch, because one must assume a very specialized form for both the Hamiltonian and the equation which determines the way in which the different fluctuations are coordinated. There is no motivation for these assumed forms except that they give the correct answer.

A fourth problem, which applies to any collapse proposal, is that in spite of a vigorous search, there is no evidence that collapse occurs. All the considerable evidence, primarily on interference effects on several-body systems (up to 700 nucleons; and perhaps a billion electrons in SQUID experiments [18]), so far points to no collapse (that is, the interference is just what is predicted by no-collapse quantum mechanics). Also, collapse should cause changes in certain decay processes but these are not observed. So there is currently no experimental reason to suppose the probability law can be explained by collapse.

In summary, collapse cannot be ruled out but the current experimental and theoretical status makes it quite unlikely (in my opinion).

## Hidden Variable Interpretations.

In hidden variable interpretations, one branch of the wave function is marked or singled out by "hidden," non-quantum mechanical variables which are not subject to experimental observation. As in the collapse case, one must remember there is essentially only one reason to hypothesize hidden variables, and that is as an explanation of the probability law. It is not needed to explain perception of only one version of reality, localization, and all our other perceptions of the physical world. And as with collapse, there is no experimental evidence for hidden variables. In addition, the Bell-Aspect experiment [11] gives some evidence against hidden variables; it shows there can be no *local* hidden variable interpretation. But it does not rule out more general hidden variable models.

### The Bohm hidden variable model.

One problem in ruling out hidden variable interpretations is that Bohm [20,21] devised a scheme which, within its bounds, works very well. For a single-particle wave function, one can derive a continuous infinity of trajectories in a three-dimensional space from the Schrödinger equation. It is then assumed that a "particle" is put on one of those trajectories. If it is further assumed there is a certain initial density of trajectories, then the probability law follows, at all times, from averaging over all the possible trajectories. If the wave function is describing N "particles" instead of just one, then the trajectories are in 3N-dimensional space. This scheme works perfectly—subject to certain objections—in giving the probability law in all non-relativistic cases where there is no creation or annihilation of particles.

But there are indeed objections.

**1.** If we have a single-particle wave function, the Bohm rule is that a "particle" is put on one and only one of the trajectories. But since there is no "source equation" linking particles with wave functions, the restriction to one and only one particle per single-particle wave function is completely arbitrary. There is nothing *in the mathematics* that prevents one from putting "particles" on two or more of the infinite number of trajectories (in the "single-particle" case); or from putting no "particle" on any of the "single-particle" trajectories. If one attempts to correct for this by using some sort of source equation linking the existence of a single particle with a single-particle wave function, the mathematics gets thrown off and the probability law no longer follows. So this objection will be extremely difficult (and probably impossible) to fix.

**2.** To illustrate the second problem, we do a spin ½ Stern-Gerlach experiment, with the "particled" version of the atomic state, detectors, and observer denoted by a dot, so the state, schematically, is

$$|\Psi\rangle = a(+)|+1/2\rangle|D+,\text{yes}\rangle|D-,\text{no}\rangle|\text{Obs perceives }+1/2\text{, but not consciously}\rangle + \quad (15)$$
$$a(-)|-1/2\bullet\rangle|D+,\text{no}\bullet\rangle|D-,\text{yes}\bullet\rangle|\text{Obs perceives }-1/2\text{, consciously}\bullet\rangle$$

That is, it is assumed that the single particled version of the observer corresponds to our conscious perceptions. The problem is this. The non-particled observer state |Obs perceives +1/2⟩ is a perfectly valid quantum state of the brain-body, just as valid as the |Obs perceives -1/2⟩ state. The question left unaddressed in the Bohm model, and in any hidden variable model, is: *Why don't the perfectly valid but non-particled quantum states of the brain-body ever correspond to our conscious perceptions?*

One might argue that the particles, rather than the wave functions, constitute the "actual reality." But there is no support for that in the mathematics. And it seems indefensible when one notes that the wave functions alone can account for almost all the quantitative and qualitative aspects of the physical world. One might suppose the particles themselves carry consciousness (a variant of panpsychism). But that is not an aspect of any standard hidden variable scheme, and it is very difficult to carry out in practice. So I don't see any way to devise a satisfactory explanation for why the non-particled quantum versions cannot correspond to our perceptions.

**3.** The density of trajectories initially assumed is very specialized. There is no indication that, starting from any reasonable initial density, the density of trajectories will "thermalize" to that specialized density. So an extremely unlikely initial condition on the density must be assumed in Bohm's model.

In conclusion, I can see no reason to be at all optimistic that there is an acceptable hidden variable interpretation of quantum mechanics.

**The recent work of Lapkiewicz, Zeilinger, et al.**

Recently Lapkiewicz, Zeilinger, et al. [22] have carried out an experiment on single photons which excludes a large class of hidden variable theories, those which are *non-contextual*. This means that the measured value of a quantity, say the z-component of spin of a spin 1 particle, is independent of the context in which the measurement is made. Unfortunately this is not sufficient to rule out all hidden variable theories, because the entirely correct (within its domain) model of Bohm is contextual (and non-local).

To briefly explain, suppose one has a spin 1 bound state of two spin 0 particles. Then Bohm's ideas can be applied to obtain a valid hidden variable model of this simple system so that each of the possible values of $s_z$ will be detected with the appropriate probability. If this were a non-contextual model, then this system would yield the same value for $s_z$ no matter what the measuring process. But in the Bohm model, the measured value of $s_z$ depends critically on *where* the measuring device is placed because of the chaotic motion of the "particle" introduced in the Bohm model. Even a change of a thousandth of a millimeter in the location of the magnet in a Stern-Gerlach experiment would, in general, change the measured output because it would change which trajectory (+1,0,-1) the particle enters into and travels on after the magnet. Thus the Bohm model is a *contextual* hidden variable model because the measured value depends on the context—where the measuring apparatus is placed—and so the L-Z work does not apply to it. So we see that the L-Z experiment does not rule out the most likely class of hidden variable theories.

## 4. Derivation of the Probability Law.

The QM-A scheme is very powerful in that, if we leave out probability, it can satisfactorily account for all our perceptions of the physical world. But the status of the probability law in relation to QM-A is not readily apparent. Is it an entirely separate law of nature, or does it follow from the basics of QM-A (linearity and the Hilbert space structure)? We will show here that if one *assumes* probability is inherent in QM-A, and if one assumes the time evolution operator is unitary, then the $|a(k)|^2$ formula for the probability law can be derived.

**Statement of the Probability Law**

The probability law can be stated in the following way, using the state vector of Eqs. (1) and (3).

> The Probability Law. If many runs of an experiment are done, "my" perceptions match the perceptions of version *k* of the observer on a fraction $|a(k)|^2$ of the runs.

Three comments: The major one is that there is no (single) "me" in QM-A; instead there are *n versions* of "me," so that the "my" in the law doesn't make sense. This inability to match a straightforward statement of the probability law with the structure of QM-A is a huge problem for that otherwise completely successful interpretive scheme. (See Sec. 5.)

The second comment is that the only feature of the state vector which probability depends on is the coefficients; it doesn't depend on any other property such as the details of the state vectors of the atomic system, the detector, the environment or the observer. This somewhat remarkable feature will follow from our derivation. Third, probability can only refer to probability of *perception* in QM-A, because all versions of reality always *exist*.

**Assumptions.**

We need to make several assumptions in order to derive the formula for probability in QM-A. The first one is incorrect, as we will see in Sec. 5, but we will ignore that problem for now and address it in Sec. 6.

Assumption 1: *Probability of perception is a valid concept in QM-A.*

Assumption 2: *The time evolution is unitary, so the sum of the squares of the coefficients a(k) is always 1.*

Assumption 3: *The choice of basis vectors cannot affect the outcome of a process. So any convenient basis can be used in a particular problem.*

Assumption 4: Based on the non-communication between orthogonal branches in QM-A, we assume *the probability of perceiving outcome k is independent of whether or not measurements are made on other outcomes.*

Assumption 5: P*robability is assumed to be a differentiable function of the coefficients.*

Assumption 6: *We do successive measurements on two different systems, with the second measurement depending on the results of the first. If the probability of outcome k for the first measurement is $P_k$, the probability of outcome j for the second measurement, given k for the first, is $P_{2,j\ given\ k}$ and the probability of outcome k for the first and j for the second is $P_{1+2,k,j}$ for the compound system, then $P_{1+2,k,j} = P_k P_{2,j\ given\ k}$*

**Derivation.**

**1 Functional dependence of the probabilities:** To start the derivation, we do an experiment in which only one of the several possible outcomes on a particle in state $|\psi\rangle$ is detected. If we have a Stern-Gerlach experiment, for example, we put a detector on only one of the paths. Or if we have a scattering experiment, we use only one grain of film instead of many thousands. We then choose the following basis for the particle state. The first basis vector, $|1\rangle$, is the one relevant to the detected outcome. The relevant coefficient for state $|1\rangle$ is

$$a(1) = \langle 1|\psi\rangle \qquad (16)$$

The second basis vector is the part of $|\psi\rangle$ orthogonal to $|1\rangle$. Properly normalized, this is

$$a(2')|2'\rangle = |\psi\rangle - a(1)|1\rangle \tag{17}$$

with $\langle 1|2'\rangle = 0$. By adjusting the phases in the definition of $|1\rangle$ and $|2'\rangle$, we can make $a(1)$ and $a(2')$ real. The other $n-2$ basis vectors are chosen to be orthogonal to $|1\rangle$ and $|2'\rangle$ and to each other. So we see from Eq. (17) that, in terms of that basis,

$$|\psi\rangle = a(1)|1\rangle + a(2')|2'\rangle \tag{18}$$

$\langle\psi|\psi\rangle = 1$ then implies $a(2')^2 = 1 - a(1)^2$ so that

$$|\psi\rangle = a(1)|1\rangle + \sqrt{1 - a(1)^2}|2'\rangle \tag{19}$$

Now consider the probability of the detection of state 1. We see from Eq. (19) that only $a(1)$ occurs, so as far as the coefficients are concerned, the probability can depend only on $a(1)$.

$$\text{Probability of perceiving state } |1\rangle = P_1(a(1), u) \tag{20}$$

That is, $P_1(1)$ is independent of the $a(2)$, $a(3)$ and so on of Eq. (1) because those coefficients don't occur in the expression for $|\psi\rangle$ in Eq. (19). Here $u$ could be anything else besides the coefficients (such as the details of the various state vectors), that the probability might depend on. We can then use assumption 4, plus the fact that the above argument could have been done for any $k$, to infer that, for all $k$,

$$\text{Probability of perceiving state } |k\rangle = P_k(a(k), u). \tag{21}$$

where each $P_k$ could be a different function. That is, the probability of perceiving result $k$ is independent of all coefficients except $a(k)$.

**2. General form of the probabilities:** The functional form of $P_k(a(k), u)$ can now be partly determined. We know from the definition of probability that

$$\sum_{k=1}^{n} P_k(a(k), u) = 1 \tag{22}$$

We also have the normalization condition

$$\sum_{k=1}^{n} |a(k)|^2 = 1 \tag{23}$$

To derive the functional form for $P_k$, suppose we vary $a(1)$ and $a(2)$ in Eq. (22) subject to the constraint of Eq. (23). The constraint implies

$$2a(1)\delta a(1) + 2a(2)\delta a(2) = 0 \tag{24}$$

The variation of Eq. (22) then gives

$$\frac{\partial P_1(1)}{\partial a(1)}\delta a(1) + \frac{\partial P_2(2)}{\partial a(2)}\delta a(2) =$$
$$\left(\frac{\partial P_1(1)}{\partial a(1)} - \frac{\partial P_2(2)}{\partial a(2)}\frac{a(1)}{a(2)}\right)\delta a(1) = 0$$

where the 0 comes from the variation of 1. Thus

$$\frac{1}{a(1)}\frac{\partial P_1(a(1),u)}{\partial a(1)} = \frac{1}{a(2)}\frac{\partial P_2(a(2),u)}{\partial a(2)} = \lambda(u)$$

$$\tag{25}$$

The first term is a function of $a(1)$ while the second term is a function of $a(2)$. This implies that $\lambda$ is independent of the $a$'s. We can solve Eq. (25) for the probabilities to get

$$P_k(a(k),u) = (\lambda(u))/2)|a(k)|^2 + c(k,u), \quad k=1,2 \tag{26}$$

Since states 1 and 2 were chosen arbitrarily, we see that the equation holds for all $k$, not just $k=1$ and 2.

Note that the derivation doesn't hold in the two-state case because $a(2)$ is a function of $a(1)$. In fact, as a counterexample to Eq. (26), we can choose $P_1=a_1^2+f(a_1^2)$, $P_2=a_2^2+f(a_2^2)$ where $f(x)$ is any odd function of $x-1/2$ with $f(0)=0$, and this choice satisfies all the constraints on probability.

**3. Auxiliary experiments.** To deal with the two-state case and also to show that all the $c(k,u)$ of Eq. (26) must be zero, we will use an idea similar to Zurek's auxiliary experiments [1,2], outlined in Sec. 7. We have an initial atomic-level state, with $n$ components, like that of Eq. (1). After the detectors detect this state, we have Eq. (2),

$$|\Psi_1\rangle = \sum_{k=1}^n a(k)|k\rangle \ |D{:}k,\text{yes}\rangle \tag{27}$$

If the set of detectors D registers a "yes" for a given $k$, we do an auxiliary experiment on a 3-state system, with a *different system* for each $k$. Leaving out the detectors, the state is

$$|\Psi_{1,2}\rangle = \sum_{k=1}^{n}\sum_{j=1}^{3} a(k)b(k,j)|k\rangle|j,k\rangle \tag{28}$$

$$\sum_{k=1}^{n}|a(k)|^2 = 1, \quad \sum_{j=1}^{3}|b(j,k)|^2 = 1 \tag{29}$$

We now have a system with $3n$ states. We apply the reasoning of step **1** (detect only one of these compound states) to this system to arrive at the result that the probability of perceiving state $|k\rangle$ *and* state $|j,k\rangle$ is $P_{k,j}(a(k)b(k,j),u_{1,2}))$. We then use step **2**, Eq. (26), to obtain the probabilities in assumption 6.

$$P_{1+2,k,j}=\lambda_{1,2}(u)|a(k)b(k,j)|^2+c_{1,2}(k,j,u) \tag{30}$$
$$P_{2,j\ given\ k}= \lambda_2(u)|b(k,j)|^2+c_2(k,j,u) \tag{31}$$

Assumption 6 then says that, for given $k,j$,

$$\lambda_{1,2}(u)\ |a(k)|^2|b(k,j)|^2+c_{1,2}(k,j,u)=$$
$$[\ P_k(a(k),\ u)]\ [\ \lambda_2(u)|b(k,j)|^2+c_2(k,j,u)] \tag{32}$$

Taking the derivative of both sides of the equation with respect to $|b(k,j)|^2$ and invoking Eq. (22) (which implies the ratio of the $\lambda$s must be 1) implies

$$P_k(a(k),\ u)=|a(k)|^2 \tag{33}$$

This holds for any system, even two-state systems. And we see that the constants in Eq. (26) (and therefore Eq. (32)) must be zero. Thus the probabilities depend only on the coefficients, in just the way expected. Under our assumptions, they cannot depend on any other properties of the state vectors.

## 5. No Probability within the QM-A Scheme.

*In reality*, there is a probability law; for many repetitions of an experiment, my perceptions correspond to outcome $k$ on a fraction $|a(k)|^2$ of the runs. The question to be considered in this section is whether the concept of a coefficient-dependent probability makes sense *within QM-A*.

### 1. No probability of perception in QM-A.
Suppose we consider a two-state case with $|a(1)|^2=.9$, $|a(2)|^2=.1$ and do 10,000 runs, with the observer perceiving all the results. Then QM-A tells us the following:

- There are $2^{10,000}$ versions of the observer. The perceptions of each version are equally valid in the sense that each of the quantum versions of the brain corresponds to a valid perception. The coefficients on each version do not affect the validity of perception in any way.
- These versions of the observer are the only 'entities' that perceive. There is no (singular) "I" that has been assigned to a particular version with some probability in QM-A, both because there is no (singular) "I" and because there is no probabilistic assigning process in QM-A.
• That is *all* QM-A tells us about perception; there is no concept of *probability* of perception, there are just $2^{10,000}$ equally valid versions of the observer, with each version perceiving its respective outcome.

### 2. Probability implies a bias, but there is none in QM-A.

This is a more specific argument than **1**. Let n(1) denote the number of 1 states perceived by a particular version of the observer in the two-state, 10,000-run case. Then the ratio of the number of observer states with n(1) near 5,000 to those with n(1) near 9,000—the result predicted by the probability law—is about $10^{800}$. The question is: For every run of 10,000, why do our perceptions *always* correspond to a result with n(1) near 9,000, where there are, relatively speaking, extremely few versions of the observer? Why don't our perceptions ever correspond to a result with n(1) near 5,000, which is associated with many, many more versions of the observer?

The answer is that obviously *in reality*—as opposed to in QM-A—there is a *bias* in choosing which of the $2^{10,000}$ versions of the observer correspond to "my" perceptions. But *in QM-A*, there is no biasing principle. *Every one* of the $2^{10,000}$ versions of "me" perceives its respective result with equal validity; no one version or set of versions of the observer is singled out in any way.

Thus QM-A alone cannot account for the perceived probabilistic bias in the large-N case.

### 3. No acceptable statement of the probability law in QM-A.

There is another, somewhat unconventional way of showing that QM-A does not support the concept of probability. The principles upon which it is based follow directly from QM-A.

> • The probability law pertains to the probability of *perception* of outcome *k*. (All outcomes simultaneously *exist* in QM-A, so probability cannot pertain to probability of existence.)
> • The only entities that perceive in QM-A are the versions of the observer, so the probability law must be couched in terms of the *perceptions of those versions*.
> • The "probability" of version *j* of the observer perceiving state *k* is $\delta_{jk}$, which is obviously independent of the coefficients.
> • The perceptions of each version are equally valid; no version is preferred over another, no version is more aware than another in QM-A.

So the question is: Can we state the probability law in terms of the perceptions of the *versions* of the observer (which are the only perceiving entities)? That is, can we state the probability law without using "the" observer, "my" perceptions, and so on, and without using the passive voice—state *k* is perceived…—which does not name the perceiver?

We can give a couple of attempts.
- The probability of version *k* perceiving state *k* is $|a(k)|^2$. But of course that is not acceptable because the probability of version *k* perceiving state *k* is 1.
- The probability of my perceptions corresponding to those of version *k* is $|a(k)|^2$. Again, this is not acceptable because of the use of "my" perceptions; there is no "me" in QM-A.

It is our contention that—(1) because the law must be given in terms of the *perceptions of the versions*, (2) because those perceptions are not probabilistic, and (3) because no "outside observer" is probabilistically assigned to just one version—there can be no satisfactory statement of the probability law within QM-A. Thus pure QM-A is not compatible with probability. Or we can conclude somewhat more cautiously that until an acceptable statement of the probability law is given, there is no justification for assuming pure QM-A is compatible with the probability law.

**Subjective likelihood.**

There have been attempts to get around the first argument above by introducing probability into QM-A through a "subjective" process [23-28]. After the experiment is done and the result recorded by the versions of the apparatus, but before the observer sees the recorded results, each version of the observer is uncertain about what he will perceive. This *uncertainty* is said to lead to a (coefficient-dependent) *probability* for *expectations* of perceptions. So let us do the experiment again, in two steps. At first we do not allow the observer to look at the list of 10,000 results. Let n(1) again denote the number of 1 states perceived by a particular version of the observer. Then the subjective likelihood approach says that each version of the observer should, within the QM-A system, have some rational expectation about the approximate value of n(1) he will see when he looks at the list. And the authors claim that, because of the uncertainty, the rational expectation for the probability—and *presumably* the actual result when he does look, although I don't believe this step from expectation to actual has been fully addressed—can be shown to be $N|a(k)|^2$.

But no reason is given for why, within QM-A (as opposed to in reality), the versions of the observer should have any coefficient-dependent expectations for what they will perceive. Further, I don't see how these authors can make the bias associated with the $N|a(k)|^2$ law consistent with QM-A. And perhaps most importantly, these authors have not given a definition of probability in QM-A which satisfies the constraints listed in the third argument above. So I see no way the subjective likelihood approach can currently be justified. (See Kent [29] for the same conclusion.)

# 6. A non-quantum mechanical aspect to the observer. The many-worlds, one-MIND interpretation.

**Motivation.**

We have seen that QM-A, in which only the state vectors exist, can explain everything except the probability law, so it is reasonable to suppose that physical reality consists of the state vectors alone—no hidden variables, no particles. In addition, we have argued that there is no reason to be optimistic that either a collapse or a hidden variable interpretation will succeed. So it is also reasonable to suppose that QM-A contains the sum total of the mathematics used to describe physical reality—no collapse, just linear operators acting in a Hilbert space. How then, under these suppositions, are we to deal with the probability law?

To get some idea, we recall our perfectly valid definition of the law:

> If many runs of an experiment are done, "my" perceptions correspond to the perceptions of version $k$ of the observer on a fraction $|a(k)|^2$ of the runs.

The "my" certainly seems to imply there is, in some sense, an actual, single-version "me" who perceives just one outcome (with a certain probability) on each run of the experiment. A single-version "me" is achieved by reduction to one version in collapse interpretations, and by "marking" just one version in a hidden variable interpretation. But there is no collapse or marking in QM-A; all versions always exist, and all give equally valid perceptions (that is, they all correspond to equally valid quantum versions of the neural firing patterns of the brain). So, because there is no way to distinguish one version as "special" *within* QM-A, this conjectured "me" that perceives just one version of reality on each run must exist outside the mathematics of QM-A.

### The non-quantum mechanical individual Mind.

We will give the proposed interpretation in two steps, first for the individual and then more generally. To start, we suppose:
(**a**) There is a non-quantum mechanical (N-QM) aspect associated with *each* observer. It is not subject to the quantum mechanical equations of motion and has no associated state vector. We call this aspect the N-QM Mind, or just the Mind.
(**b**) This Mind "lives in" or perceives in just *one* of the universes or versions of reality (branches) allowed by quantum mechanics. Within that single universe, it perceives only the state vector of the observer's brain-body.
(**c**) The Mind is the basic source of our "conscious" perceptions. When I say "my" perceptions, I refer, at base, to the perceptions of my associated Mind (with the perceptions greatly facilitated, of course, by the mechanisms of the quantum brain-body).
(**d**) It is this "me," the N-QM Mind, perceiving just one version of reality, that the probability law applies to in Sec. 4. It is how we give validity to the first assumption there; the probability of perceiving state $k$ refers to the probability of the (single-version) N-QM Mind perceiving the quantum brain state corresponding to |Obs perceives $k$ $\rangle$.

### The need for a linkage between individual N-QM Minds.

There is, however, a problem with the *individual* N-QM Mind proposal. Suppose two observers observe a two-state system. The state vector is

$$|\Psi\rangle = a(1)|1\rangle|\text{Det reads 1}\rangle|\text{Obs } a \text{ perceives 1}\rangle|\text{Obs } b \text{ perceives 1}\rangle + \quad (34)$$
$$a(2)|2\rangle|\text{Det reads 2}\rangle|\text{Obs } a \text{ perceives 2}\rangle|\text{Obs } b \text{ perceives 2}\rangle$$

Suppose Mind $a$ perceives the state 1 quantum version of observer $a$. The quantum mechanical aspect of observer $a$ can only communicate with the quantum mechanical aspect of observer $b$ on the same state 1 branch. So it must be, if we don't want "aware" versions communicating with "non-aware" versions ("mindless hulks" [30]), that Mind $b$ also perceives state 1. That is, the different Minds must agree on the single quantum version of reality they perceive. This implies there must be a linkage between the Minds of different individuals. The most straightforward scheme for implementing this is to suppose the following.

**The over-arching MIND.**
**(A)** There is a single over-arching MIND that has no associated state vector and is not subject to the equations of quantum mechanics.
**(B)** The MIND perceives the state vector of just one branch of the full state vector of the physical universe.
**(C)** On each branch, there is a quantum mechanical version of the brain-body of each individual "self." The *individual* N-QM Mind is the part or aspect or facet of the over-arching N-QM MIND that perceives the state vector associated with the brain-body of that individual "self." Each individual Mind is therefore limited to perceiving in only a small part (one brain-body's worth) of the Hilbert space of one branch.

As in **(c)**, the Mind is the basic source of my "conscious" perceptions. When I say "my" perceptions, I refer, at base, to the perceptions of my associated Mind.

Note that the individual Minds—which are just limited-perception aspects of the over-arching MIND—perceive versions of their respective brain-body state vectors which are *on the same branch*—the branch perceived by the MIND. Thus there is now automatic agreement among the individual Minds on the singled-out version of "reality." (No more talking to mindless hulks.)
**(D)** It is now both the MIND and the individual Minds, perceiving just one version of reality, which the probability law applies to in Sec. 4. It is how we give validity to the first assumption there, with m-w* (see below) substituted for QM-A; the probability of perceiving state $k$ refers to the probability of the MIND and relevant Minds perceiving the quantum state corresponding to $|\text{Obs perceives } k\rangle$.

**The many-worlds, one-MIND or m-w* interpretation.**
We call this scheme—the QM-A mathematics plus assumptions **(A)** through **(D)**—the many-worlds, one-MIND, or m-w*, interpretation. We see from Secs. 2 and 4 that this interpretation can account for all phenomena, including the probability law.

**Comments.**
**1.** Since the N-QM aspects are outside the mathematics of quantum mechanics, we do not understand the act of perception in m-w*. But it almost certainly does not make use of photons. The fact that the N-QM MIND, outside the laws of quantum mechanics, perceives the state vector, which is governed by the laws of quantum mechanics, is perhaps the weakest link in this interpretation.

**2.** This is not a collapse scheme. The MIND does not collapse the wave function (nor do the Minds); it simply perceives or concentrates its awareness on just one branch of the state vector. The very successful mathematics of QM-A is not altered or amended in any way.

**3.** Living beings are different from non-living objects in this interpretation. Detectors, rocks, and so on do not have N-QM Minds associated with them. They consist only of state vectors governed entirely by quantum mechanics.

**4.** Do we really need to take on this metaphysical (or supra-physical) baggage? At this point, we are simply exploring how much baggage is necessary. That is, we are exploring what assumptions need to be made to obtain an interpretation which includes the probability law (1) if physical reality consists solely of the state vectors and (2) if QM-A is the only mathematics allowed.

Admittedly, assuming there is something essential outside the mathematics is a somewhat radical proposal. However, if QM-A describes physical reality so well, and if it is found that hidden variables and collapse schemes don't work (and they haven't succeeded in three quarters of a century of effort), what are the alternatives? There are certainly other possible interpretations, but none of them, it seems to me, gives sufficient weight to the astonishing successes of the QM-A scheme. When the state vectors accurately describe the hydrogen atom, spin ½, Bell-like experiments on entangled systems, the consequences of internal symmetries, and a myriad other systems so well, and when there is no physical system wrongly described, it is hard to escape the conclusion that physical reality is constructed solely from state vectors. If that is true, and if there is no collapse or hidden variables, then one is inexorably led to m-w*.

Also, to put the one-MIND proposal in perspective, it is not a trivial accomplishment that m-w* explains all of our perceptions, including perception of a single version of reality, the probability law, and agreement among observers with no alterations of or amendments to the very successful mathematics of quantum mechanics. And the assumptions are relatively few and simple.

**5.** Is the m-w* interpretation experimentally testable? It might be, through a test of the probability law. To illustrate, let us temporarily ignore the results of Sec. 4 and imagine for a moment that the perceptions of the MIND are governed by a "probability" law,

$$P(|a(k)|^2) = .8|a(k)|^2 + .2|a(k)|^4 \tag{35}$$

for perception of a state with amplitude $a(k)$. Now suppose we repeat an experiment N times, with N very large, for a two-state system. Then the amplitude squared relevant to perception of $n$ "1" results at the end of this process is

$$|A(n)|^2 = N!/(n!(N-n)!) |a(1)|^{2n} |a(2)|^{2(N-n)} \tag{36}$$

If we put this amplitude squared into Eq. (35) for P, we see that, because $A^2(n)$ has a sharp maximum as a function of $n$ at $n/N=|a(1)|^2$, there will also be a sharp maximum in the *probability* function for the N-run experiment at $n/N=|a(1)|^2$, so $n$ will always be perceived as being near $N|a(1)|^2$. That is, the probability law "in the large," Born's rule, is pretty much independent of the "probability law" "in the small." (Of course this could not happen in a classical scheme.) In fact, one need not even have a well-defined probability law in the small at all so long as states with much larger amplitudes are much more likely to be perceived (by the MIND and the individual Minds).

So an experimental test of m-w*, although not a conclusive one, is to have the observer perceive the outcome of every run of the experiment, not just the end result after a large number of repetitions. It is possible that one would observe something other than the usual $|a(k)|^2$ law in that case.

This reasoning may give some insight into the "cause" of the probability law. We suppose the MIND is very likely to perceive states with relatively large amplitudes, almost as if amplitude corresponded to "brightness." Then the $|a(k)|^2$ probability law arises, not because of a random "mathematical" process, but because of the amplitude of Eq. (36) and the propensity of the MIND to see large-amplitude states.

**6.** This interpretation is in the same ballpark as Albert and Loewer's many-minds version [30,31] of Everett's scheme in that it has the same motivation; QM-A works extremely well so we should not change the QM-A mathematics. But to get around the mindless-hulk problem, they suppose each sentient being is associated with a continuous infinity of minds; whereas we suppose there is a single over-arching MIND and each individual Mind is a facet of the MIND. Albert and Loewer assign probability to the distribution of the continuous infinity of minds whereas we presume it arises from the perception of large-amplitude states by the MIND.

**7.** What about choice? We will not address that question here except to say that the individual N-QM Mind almost certainly does not have a choice of whether to see a live cat or a dead cat. (This point may have a tie-in with comment 3.)

## 7. Zurek's Derivation of the Probability Law.

Zurek claims to have derived the probability law from what is essentially QM-A. But I believe his derivation rests on two unwarranted assumptions. First, he *assumes* QM-A includes probability (of perception). This is stated in his Fact 2 in references [1] and [2]: "Given the measured observable, the state of the system *S* is all that is needed … to predict measurement results, including probabilities of outcomes." This assumption, however, is contradicted by the reasoning of our Sec. 5, which shows there is no *a priori* concept of probability in the pure linear equation, Hilbert space quantum mechanics of QM-A. Note that it is not appropriate to reason that, because probability occurs in Nature, it must be contained in QM-A. If Zurek's reasoning is to stand, he must show *probability is inherent in QM-A*. That is, he must refute the arguments of Sec. 5. In particular, he must give a statement of the probability law which is acceptable within the confines of QM-A (Sec. 5.3).

Second, to derive the probability law, Zurek uses an ingenious method involving ancillary experiments. But I think there is an unwarranted assumption here also. To illustrate, suppose we consider the spin ½ state

$$|\Psi_0\rangle = \sqrt{(3/5)}\,|1\rangle + \sqrt{(2/5)}\,|2\rangle \qquad (37)$$

and we do a Stern-Gerlach experiment on it. There is a detector on each of the two possible paths, so after detection the state is

$$|\Psi_1\rangle = \sqrt{(3/5)}\,|1\rangle|d,1{:}\text{yes}\rangle|d,2{:}\text{no}\rangle + \qquad (38)$$
$$\sqrt{(2/5)}\,|2\rangle|d,1{:}\text{no}\rangle|d,2{:}\text{yes}\rangle$$

Now if detector 1 reads yes, we do an ancillary Stern-Gerlach experiment on a spin 1 system (denoted by a single prime), with equal coefficients ($1/\sqrt{3}$), and if detector 2 reads yes, we do an ancillary Stern-Gerlach experiment on another spin ½ system (denoted by a double prime), with equal coefficients ($1/\sqrt{2}$). Then, taking into account the multiplication of the coefficients, and leaving out the detector states, the final state is

$$|\Psi_2\rangle = \sqrt{1/5}\,\big[|1\rangle|1'\rangle + |1\rangle|2'\rangle + |1\rangle|3'\rangle + |2\rangle|1''\rangle + |2\rangle|2''\rangle\big] \qquad (39)$$

Zurek claims the equal-coefficient reasoning should apply here so all the states are equally probable (and he goes on to derive the Born rule from this). But Zurek's equal-coefficient-implies-equal-probability reasoning is based on exchanging states. If we exchange the third and fourth set of states in Eq. (39), however, we are exchanging a spin 1 state, $|3'\rangle$, with a spin ½ state, $|1''\rangle$. But an exchange of such dissimilar states is, to the best of my knowledge, not permitted in quantum mechanics. (There may be a way around this problem; use two spin 1 ancillary states but in the second, use only two of the components. But one must still justify the steps.) Perhaps even more to the point, states $|1\rangle$ and $|2\rangle$ from the original spin ½ system have been exchanged in the third and fourth states of Eq. (39) but they have not been exchanged in the first, second and fifth. This seems inconsistent. Thus I don't believe Zurek's equal-coefficient reasoning holds in this case, and that calls his derivation into question.

In conclusion, if Zurek's derivation of probability is to be considered valid, he needs to more fully justify these two assumptions.

## 8. Summary and conclusion.

The basic mathematical structure of quantum mechanics, QM-A, consists of its linearity and the Hilbert space structure of its state vectors. The problem is that the linearity leads to the simultaneous existence in the mathematics of several potential versions of reality, each equally valid—Schrödinger's cat is both alive and dead at the same time. So the first question is, in light of these multiple versions of reality, how are we to understand our perception of a single version of reality?

The most logical starting point is to carefully examine QM-A, by itself, with no assumption of collapse or the existence of particles or hidden variables, to see whether it can indeed account for this seeming gap between the mathematics and our perception. It turns out that QM-A alone can explain our only-one perceptions. But further—surprising in view of the many interpretations proposed—QM-A alone can explain almost *all* our perceptions. That is, in addition to quantitative results such as the hydrogen atom spectrum, it can explain more qualitative results including localized effects from a spread-out wave function, the photoelectric and Compton effects, the particle-like properties of mass, energy, momentum, spin and charge, wave-particle duality, and long-range correlations in Bell-type experiments. Thus as far as all these properties are concerned, there is no need for the concept of particles, for collapse, or for hidden variables.

The single exception to the successes of QM-A is the probability law; there can be no *probability* of perception in straight QM-A because each of the $n$ simultaneously existing versions of reality is equally valid; all $n$ outcomes exist and are perceived on every run. Collapse and hidden variable interpretations have been proposed as a way of dealing with this problem. However, there is no evidence for either of these interpretations and there are severe theoretical hurdles to, and experimental restrictions on, the construction of a successful collapse or hidden variable interpretation.

Because of the near-perfect success of QM-A and the lack of success for mathematical amendments to that scheme, we are led to seek an interpretation in which (1) physical reality is constructed solely from state vectors and (2) the mathematics consists solely of the linear equations of QM-A, with no collapse or hidden variables.

The primary clue about how to proceed under these assumptions comes from the probability law itself. If "I" am to perceive a single outcome with such and such a probability, then instead of there just being $n$ quantum *versions* of me that each perceive a different outcome, there must also be an aspect of me—a Mind—that is aware of only a *single* outcome on each run; that is, there must indeed be a unique "I." So to obtain a scheme that can accommodate probability as well as all the other perceived aspects of our physical world—without altering or amending the mathematics—we must suppose that each observer has an aspect, *outside the laws of quantum mechanics*, which is aware of just one quantum version of the observer's brain-body on each run of an experiment; it does not collapse the wave function, it just *perceives* (part of) a single branch of the wave function. Probability then refers to the probability of the non-quantum mechanical aspect of the observer perceiving a certain outcome.

Once we have introduced the potential for probability into the scheme, the specific $|a(k)|^2$ form of the probability law can then be derived. It is primarily a consequence of the normalization condition $\sum |a(k)|^2 = 1$.

But the introduction of a non-quantum mechanical aspect to each observer raises a further problem. For each observer, one quantum version of the observer is singled out as the "aware" one. However, a reasonable requirement on an interpretation would be that the aware version of one observer should only be able to communicate with the aware version of another observer. For this to hold, the aware versions of different observers must agree on the single quantum version they are aware of. This implies there must be a linkage between the non-quantum mechanical selves of different individuals.

We are therefore led to suppose first, that there is, outside the mathematics of quantum mechanics, an over-arching MIND which perceives just one branch of the state vector; and second, that corresponding to each of us there is an individual Mind, also outside the mathematics of quantum mechanics, which is an aspect or facet of the over-arching MIND. This aspect perceives only that part of the state vectors corresponding to our own individual brain-body.

So we see that this m-w* interpretation, consisting of the mathematics of QM-A, the overarching MIND, and the individual Minds, can explain all our perceptions; it can explain perception of only one version of reality, the probability law, agreement among observers, and all the other quantitative and qualitative aspects of the physical world. The principles used in the derivation of these results are linearity, invariance properties, orthogonality, and probability. These are common to all forms of quantum mechanics—quantum field theory, string theory and so on—so the conclusions apply to all forms of the theory.

What are the odds that this interpretation, which goes beyond the conventional understanding of physical theory, is correct? The overriding factor in my opinion is the incredible success of QM-A alone in describing our perceptions; it's only failure is that it cannot accommodate the probability law. That fact strongly suggests that the unamended QM-A should be the sole mathematical component of any interpretation, and that the state vectors should be the sole physical component. A second factor is that the difficulties facing other interpretations have not been resolved, even after 76 years (since Einstein, Podolsky, Rosen [32] and Schrödinger's cat [33]) of effort by the best minds in physics. Thus one is inclined to consider the many-worlds, one-MIND scheme as the most likely interpretation at the present time.

## References

[1] Wojciech Hubert Zurek, *Relative states and the environment: einselection, envariance, quantum Darwinism, and the existential interpretation*, arXiv:quant-ph/0707.2832v1 (2008).
[2] Wojciech Hubert Zurek, *Probabilities from entanglement, Born's rule from envariance,* arXiv:quant-ph/0405161v2 (2005).
[3] Hugh Everett, III, *Relative state formulation of quantum mechanics.* Rev. Mod. Phys. 29 454 (1957).
[4] Casey Blood, *No Evidence for particles,* arXiv:quant-ph/0807.3930v2 (2011).
[5] Casey Blood, *Constraints on Interpretations of quantum mechanics*, arXiv:quant-ph/0912.2985 (2009).
[6] H. Dieter Zeh, *Decoherence: Basic Concepts and Their Interpretation*, arXiv:quant-ph/9506020v3 (2002).
[7] H. Georgi and S. L. Glashow, *Unity of all elementary-particle forces.* Phys. Rev. Lett., 32, 438 (1974).
[8] Robert E. Marshak, *Conceptual Foundations of Modern Particle Physics* (World Scientific, Singapore, 1993).
[9] E. P. Wigner, *On Unitary Representations of the Inhomogeneous Lorentz Group*, Ann. Math. 40, 1 (1939).
[10] J. S. Bell, *On the Einstein Podolsky Rosen paradox*, Physics, 1, 195 (1964).

[11] A. Aspect, P. Grangier, and G. Rogers, *Experimental Realization of Einstein-Podolsky-Rosen-Bohm Gedankenexperiment: A New Violation of Bell's Inequalities*, Phys. Rev. Lett. 47, 460 (1981).
[12] V. Jacques *et al*, *Experimental Realization of Wheeler's delayed-choice Gedankenexperiment*, Science 315, 966 (2007).
[13] M. O. Scully and K. Druhl, *Quantum eraser: A proposed photon correlation experiment concerning observation and "delayed choice" in quantum mechanics* Phys. Rev. A 25, 2208 (1982).
[14] S. P. Walborn, M. O. Terra Cunha, S. Paua, and C. H. Monken, *Double Slit Quantum Eraser*, Phys. Rev. A, 65 033818, (2002).
[15] G. C. Ghirardi, A. Rimini, and T. Weber, *Unified dynamics for microscopic and macroscopic systems*, Phys. Rev. D34, 470 (1986); Phys. Rev.D36, 3287 (1987).
[16] Philip Pearle, *How Stands Collapse I*, arXiv:quant-ph/0611211v1 (2006).
[17] Philip Pearle, *How stands collapse II.* arXiv:quant-ph/0611212v3 (2007).
[18] A. J. Leggett, *Testing the limits of quantum mechanics: motivation, state of play, prospects,* Journal of Physics: Condensed Matter 14 R415-R451 (2002).
[19] Roger Penrose, *The Road to Reality*, (Vintage Books, New York, 2007).
[20] David Bohm, *A suggested interpretation of quantum theory in terms of "hidden variables,* Phys. Rev. 85 166,180 (1952).
[21] D. Bohm and B. J. Hiley *The Undivided Universe* (Routledge, New York, **1993).**
[22] Radek Lapkiewicz, Anton Zeilinger, et al, *Experimental non-classicality of an individible quantum system*, Nature, June 23, 2011 | DOI: 10.1038/nature10119
[23] Lev Vaidman, *On schizophrenic experiences of the neutron or why we should believe in the many-worlds interpretation of quantum theory.* International Studies in the Philosophy of Science, Vol. 12, No. 3, 245-261 (1998).
[24] Lev Vaidman, *Many-Worlds Interpretation of Quantum Mechanics.* Stanford Encyclopedia of Philosophy. (http://plato.stanford.edu/entries/qm-manyworlds/) (2002).
[25] David Deutsch, *Quantum Theory of Probability and Decisions*, Proceedings of the Royal Society of London A455 3129-3137 (1999).
59
[26] David Wallace, *Quantum probability from subjective likelihood: improving on Deutsch's proof of the probability rule,* arXiv:quant-ph/0312157v2 (2005).
[27] Simon Saunders and David Wallace, *Branching and uncertainty*, British Journal for the Philosophy of Science, 59(3):293-305 (2008).
[28] David Wallace, *The quantum measurement problem: state of play*, arXiv:quant-ph/0712.0149v1 (2007).
[29] Adrian Kent, *One world versus many: the inadequacy of Everettian accounts of evolution, probability, and scientific confirmation*, arXiv:0905.0624v2 (2010).
[30] David Z. Albert, *Quantum Mechanics and Experience*, Harvard Univ. Press, Cambridge, MA, (1994).
[31] D. Albert and B. Loewer, *Interpreting the many-worlds interpretation*, Synthese 77, 195 (1988).

[32] A. Einstein, B. Podolsky, and N. Rosen, *Can quantum mechanical description of physical reality be considered complete?* Physical Review, 47, 777-780, (1935).
[33] Erwin Schrödinger, *The Present Situation in Quantum Mechanics,* Die Wissenschaften, 23, 807-812, 824-828, 844-849, (1935).